\documentclass[preprint,aps,superscriptaddress]{revtex4-1}
\usepackage{amssymb}
\usepackage{amsfonts}
\usepackage{amsmath}
\usepackage{amsxtra}
\usepackage{amscd}
\usepackage{amsthm}
\usepackage{graphicx}
\usepackage{color}
\usepackage{epsf}
\usepackage{bbold}
\usepackage{xcolor}
\usepackage{mathalfa}
\usepackage{eucal}

\renewcommand{\>}{{\rangle}}
\newcommand{\eq}[1]{Eq.(\ref{#1})}
\newcommand{\eps}{\epsilon}

\begin{document}

\title{Films of rhombohedral graphite as two-dimensional topological semimetals}

\author{Sergey Slizovskiy}
\affiliation{National Graphene Institute, The University of Manchester, Booth Street East, Manchester, M13 9PL, UK}
\affiliation{Department of Physics \& Astronomy, The University of Manchester, Oxford Rd., Manchester, M13 9PL, UK}
\affiliation{NRC ``Kurchatov Institute'', St. Petersburg INP, Gatchina, 188 300, St. Petersburg, Russia}

\author{Edward McCann}
\affiliation{Physics Department, Lancaster University, Lancaster, LA1 4YB, UK}

\author{Mikito Koshino}
\affiliation{Department of Physics, Osaka University, Toyonaka 560-0043, Japan}

\author{Vladimir I. Fal'ko}
\affiliation{National Graphene Institute, The University of Manchester, Booth Street East, Manchester, M13 9PL, UK}
\affiliation{Department of Physics \& Astronomy, The University of Manchester, Oxford Rd., Manchester, M13 9PL, UK} 
\affiliation{Henry Royce Institute for Advanced Materials, Manchester, M13 9PL, UK}

\begin{abstract}
Topologically non-trivial states appear in a number of materials ranging from spin-orbit-coupling driven topological insulators to graphene. In multivalley conductors, such as mono- and bilayer graphene, despite a zero total Chern number for the entire Brillouin zone, Berry curvature with different signs concentrated in different valleys can affect the material's transport characteristics. Here we consider thin films of rhombohedral graphite, which appear to retain truly two-dimensional properties up to tens of layers of thickness and host two-dimensional electron states with a large Berry curvature, accompanied by a giant intrinsic magnetic moment carried by electrons. The size of Berry curvature and magnetization in the vicinity of each valley can be controlled by electrostatic gating leading to a tuneable anomalous Hall effect and a peculiar structure of the two-dimensional Landau level spectrum.
\end{abstract}

\maketitle

{\bf Introduction}

Berry curvature~\cite{berry84,xiao10} is a measure of the topological nature of non-trivial states appearing in materials ranging from spin-orbit-coupling driven topological insulators~\cite{konig07,hsieh08,hasan10,asboth16} to  graphene~\cite{novo05,zhang05,novo06}. In multivalley conductors, such as graphene, Berry curvature with different signs concentrated in different valleys~\cite{xiao07} can affect the material's observable properties
even though the Chern number for the entire Brillouin zone may be zero.
Recently, there has been renewed interest in rhombohedral graphite~\cite{mcclure69} due to progress in fabricating and characterizing thin films~\cite{pierucci15,henni16,henck18,myhro18,lat19,geisenhof19}.
Rhombohedral is one of the structural phases of graphite which has a specific `ABC' stacking of consecutive honeycomb layers of carbon atoms such that every atom has a nearest neighbor from an adjacent layer either directly above or underneath it. The interlayer hybridization of all carbon $P_z$ orbitals with characteristic energy  $\gamma_1 \approx 0.38$ eV leads to a gapped electronic spectrum in the middle of a thin film of $N$ layers (the bulk gap $\approx 3\pi \gamma_1/N$, see Methods). However, in the surface layers of the film, half of the carbon atoms don't have a nearest neighbor in the next layer for hybridizing their $P_z$ orbitals, leading, as in the Su-Schrieffer-Heeger model~\cite{su79,asboth16}, to low-energy surface states. These low-energy states form bands that cover the entire energy range within the thin film bulk gap~\cite{guinea06,manes07,min08,koshino09,heik11,xiao11,kopnin13,ho16}. 

In this article, we theoretically model thin films of rhombohedral graphite and find that they retain their two-dimensional nature for tens of layers of thickness. The low-energy surface states give rise to a semi-metallic band structure with two bands that are almost degenerate near the valley center, but split apart at $p \sim p_{\mathrm{c}}$ where the dispersion is highly anisotropic ($p_{\mathrm{c}} = \gamma_1 / v$ where $v$ is the Dirac velocity of electrons in graphene determined by the intralayer carbon-carbon hopping parameter $\gamma_0$).
The presence of spatial asymmetry between the surface layers, which may be controlled using an electric displacement field applied perpendicularly to the film~\cite{mcc06,mcc06b} or which may arise from interaction-induced spontaneous symmetry breaking~\cite{min08b,jung11b}, can create an insulator with an energy band gap and topologically non-trivial states represented by a giant Berry curvature and intrinsic magnetic moment of electrons.
We predict that the topological nature of the surface states will be manifest in electronic transport properties including a large anomalous Hall effect and anomalous transverse photoconductivity.
In addition, these features are reflected in the electronic spectra in the presence of a perpendicular magnetic field (the Landau level spectra) whereupon spatial asymmetry breaks valley degeneracy with different patterns of level crossing and hybridization in the two valleys.

\begin{figure}[!t]
\includegraphics[scale=0.49]{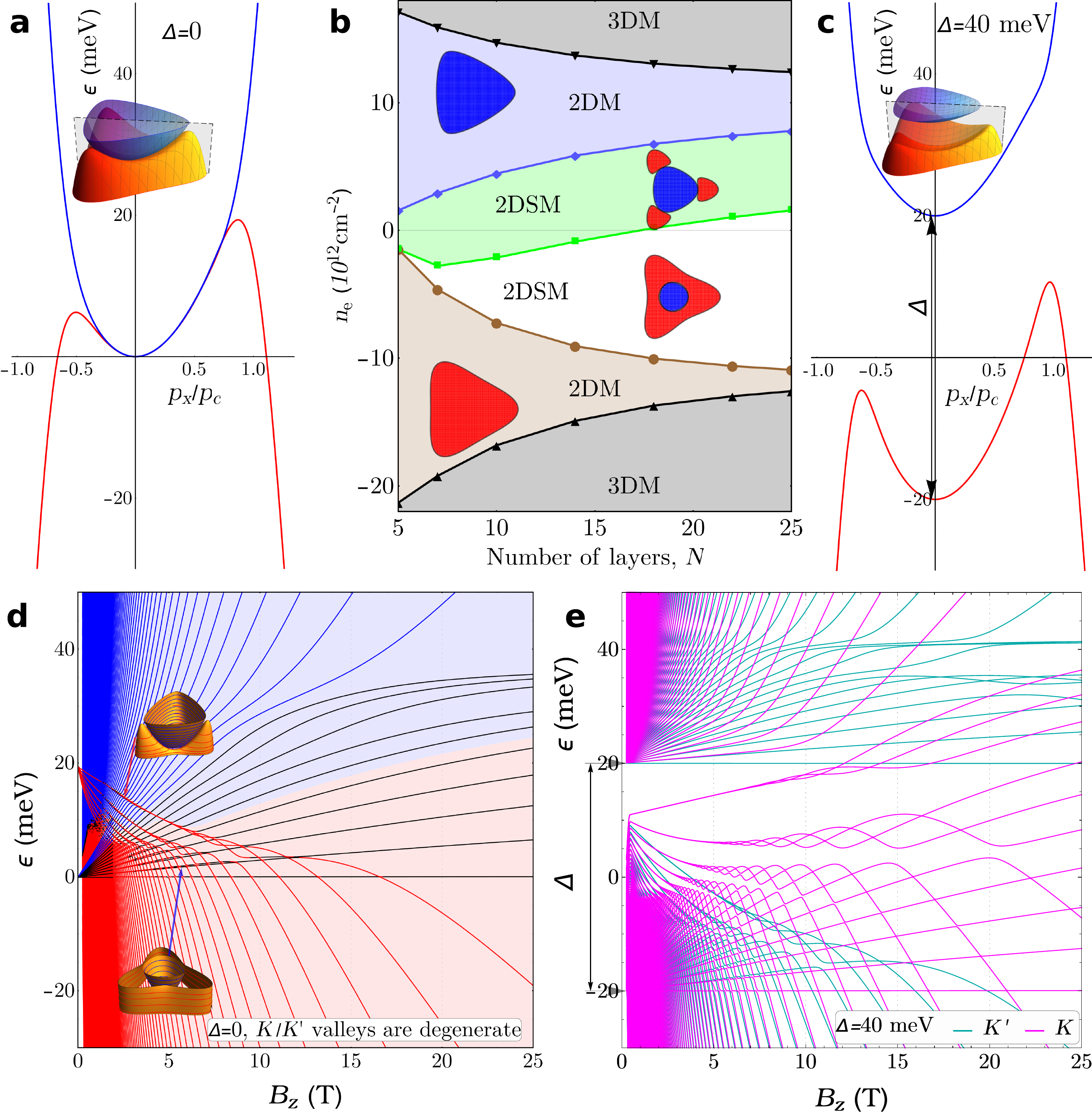}
\caption{{\bf Electron dispersion and Landau levels in a film  of rhombohedral graphite.}
Low-energy dispersion $\epsilon (p_x,0)$ as a function of the in-plane momentum $p_x$ in the $K$ valley for $N = 10$ layers with {\bf a}, top/bottom asymmetry $\Delta = 0$ and {\bf c}, $\Delta = 40\,$meV. Insets: 3D plots of $\epsilon ({\bf p})$ with parametric regimes for different Fermi surface topology shown in {\bf b} as a function of carrier density $n_{\mathrm{e}}$ and number of layers $N$ for $\Delta = 0$ with a two-dimensional semi-metal (2DSM), a two-dimensional metal (2DM) and a bulk metal (3DM). Landau level (LL) spectrum as a function of perpendicular magnetic field $B_z$ in a ten-layer film with: {\bf d}, $\Delta = 0$ and {\bf e}, $\Delta = 40\,$meV. In {\bf e}, magenta/cyan lines correspond to the $K'$/$K$ valleys. Plots were obtained by numerical diagonalization of the full hybrid $\mathbf{k} \cdot \mathbf{p}$ tight-binding model ($\hat H_N$ in Methods). In {\bf d}, red/black lines show extrapolation of LLs to $B_z \rightarrow 0$ estimated using semiclassical quantization near band edges.
}\label{fig1}
\end{figure}

\vspace{1.0cm}
{\bf Results}

{\bf Semi-metallic band structure.}
When studied taking into account all details of intra- and interlayer carbon-carbon couplings (see Methods) in the full Slonzewski-Weiss-McClure (SWMcC) tight-binding model~\cite{mcclure69,dressel02,koshino09,kopnin13}, the surface states in a thin film of rhombohedral graphite produce a semi-metallic band structure illustrated in Fig.~\ref{fig1}a.
The dispersion in Fig.~\ref{fig1}a is plotted as a function of the in-plane momentum $\mathbf{p}$ counted in a zig-zag direction from the center of the valley $K$ and normalized by $p_{\mathrm{c}} = \gamma_1 / v$.
In contrast to some earlier studies~\cite{ho16,henni16}, the dispersion in Fig.~\ref{fig1} takes into account both skew interlayer couplings $\gamma_3$ and $\gamma_4$.
The two bands of the surface states in an ABC graphite film, below referred to as conduction (blue) and valence (red), are almost degenerate near the valley center, splitting apart in the momentum range $p \agt p_{\mathrm{c}}$.
The electron dispersion at $p \sim p_{\mathrm{c}}$ is highly anisotropic, due to trigonal warping effects generated by skew interlayer hoppings, with inverted orientation  in valley~$K'$.

The interplay between these factors makes an undoped film of rhombohedral graphite a two-dimensional semi-metal (2DSM) as its Fermi level lies within both the conduction and valence surface bands. It becomes a two-dimensional metal (2DM) upon n- or p-doping when the Fermi level lies in only one of the conduction or valence surface bands and, eventually, a bulk metal (3DM) where the Fermi level lies within the bulk bands.
In Fig.~\ref{fig1}b, we identify parametric regimes for each of these three cases, taking into account the dependence of the spectrum on the number of layers.
The parametric diagram in Fig.~\ref{fig1}b was built by brute-force diagonalization of a hybrid $\mathbf{k} \cdot \mathbf{p}$ tight-binding approach model (HkpTB) based on the full SWMcC model, in which the intralayer hopping of electrons between carbon atoms is taken into account in a continuous description of sublattice Bloch states using $\mathbf{k} \cdot \mathbf{p}$ theory in the $K$ and $K'$ valleys, combined with interlayer hoppings introduced in the spirit of a tight-binding model (see Methods).
For a film with $N$ layers, this involves finding eigenvalues and eigenstates of a $2N \times 2N$ matrix acting in the space of  the sublattice Bloch states in each valley.
When studied in the presence of a vertical electric displacement field (perpendicular to the thin film), the bands responsible for the semimetallicity split by $\Delta = U_1 - U_N$, where $U_1$ and $U_N$ are the onsite energies of the surface layers, so that the system may be tuned into a gap-full insulator, Fig.~\ref{fig1}c. It may also be possible to induce a spectral gap by superconductivity~\cite{kopnin13} or spontaneous symmetry breaking into a magnetic state~\cite{zhang11,jia13,pamuk17}, making $\Delta$ spin and/or valley dependent as in the layer antiferromagnetic configurations discussed in the context of bilayer graphene~\cite{min08b,jung11b} and experimentally observed in rhombohedral graphite for $N=3$~\cite{bao11,lee14} and $N=4$~\cite{myhro18}.

When subjected to an external magnetic field perpendicular to the plane of a film, the low-energy spectrum splits into interweaving electron-like and hole-like Landau levels (LLs). For a film of rhombohedral graphite, a representative example is shown in Fig.~\ref{fig1}d computed for $N=10$ layers (see Methods), with electron-like LLs dispersing upwards for energy $E \agt 20\,$meV, hole-like LLs dispersing downward for $E \alt 0\,$meV, and both kinds of levels present in the semimetallic range $0 \alt E \alt 20\,$meV, where one can relate the electron-like and hole-like LLs to different parts of the 2D electron Fermi lines illustrated in the insets.
In Fig.~\ref{fig1}e, we show how the LL spectrum is modified by the opening of a gap (electrostatically controlled, or induced by exchange energy) which enhances some of the avoided crossings in the LL spectrum.

{\bf Two-band model.}
To develop intuition about the LL spectra shown in Fig.~\ref{fig1} as well as to anticipate the magnetoconductivity  of  an ABC film, we use a simplified two-band model which we derived from the full HkpTB by projecting the $2N \times 2N$ Hamiltonian onto the pair of surface states in the outermost (bottom and top) layers,
\begin{eqnarray}
\hat H &=& \left(
          \begin{array}{cc}
            \frac{\kappa^{\dagger}\kappa}{2m_{\star}} + \frac{\tilde \Delta }{2} & -\gamma_1 X(\mathbf{p})  \\
            -\gamma_1 X^{\dagger}(\mathbf{p}) & \frac{\kappa\kappa^{\dagger}}{2m_{\star}} - \frac{\tilde \Delta }{2} , \\
          \end{array}
        \right) \equiv \frac{p^2}{2m_{\star}} + {\bf d} \cdot {\boldsymbol\sigma},  \label{eq1} \\
\tilde \Delta &=& \Delta - \frac{e a_z \beta_1 v^2 }{\gamma_1} [ \mathbf{p} \times \mathbf{B}_{\parallel} ]_z , \
\kappa = \xi p_x + i p_y + i e x B_z/\hbar .\nonumber
\end{eqnarray}
Here, $m_{\star} \sim 0.4 m_0$ (see Methods), we use a vector ${\boldsymbol\sigma} = (\sigma_x , \sigma_y , \sigma_z)$ of Pauli matrices acting in the space of surface states and ${\bf d} = (d_x , d_y , d_z)$, and incorporate the arbitrarily-oriented magnetic field with components both perpendicular, $B_z$, and parallel, $\mathbf{B}_{\parallel} = (B_x , B_y)$, to the  film. In Eq.~(\ref{eq1}), the two valleys of graphite ${\bf K}=(4\pi / 3a , 0)$ and ${\bf K'}=-{\bf K}$ are indexed by $\xi =  1$ and $-1$ respectively. In the expression for $\tilde \Delta$ we take into account the effect of the Lorentz boost experienced by electrons tunneling between surface states,
where $\beta_1 = \frac{2\gamma_4}{\gamma_0} (2N-3) + \frac{2\delta}{\gamma_1} (N-1)$ ($\beta_1 \propto N$ for $N \gg 1$), $a_z$ is the interlayer distance and $-e$  is the electronic charge.
Parameter $\delta$ represents a difference in energy between the dimer sites inside the crystal and non-dimer sites $A_1$ and $B_N$ in the outer layers.
In the definition of operator $\kappa$ and its Hermitian conjugate, $\kappa^{\dagger}$, we use the Landau gauge for the out-of-plane magnetic field $B_z$, whereas, for $B_z=0$, $\kappa = \xi p_x + ip_y$. The off-diagonal element, $X(\mathbf{p}) = \sum \frac{(n_1+n_2+n_3)!}{n_1!n_2!n_3!}
\left(-\frac{\kappa^{\dagger}}{p_{\mathrm{c}}}\right)^{n_1}
\left( -\frac{\gamma_3}{\gamma_0} \frac{\kappa}{p_{\mathrm{c}}} \right)^{n_2}
\left( -\frac{\gamma_2}{2\gamma_1} \right)^{n_3}$,
arises from interlayer hoppings (skew hopping between neighboring layers $\gamma_3$ and `vertical' next-layer hopping $\gamma_2$ in addition to $\gamma_1$ mentioned earlier) passing electrons from one surface to another~\cite{koshino09}, and the summation is taken over integers $n_i \geq 0$ such that $n_1 + 2n_2 + 3n_3 = N$.

Qualitatively, the spectrum of $\hat H$, $\epsilon_{\pm} ({\bf p}) = p^2/(2m_{\star}) \pm |{\bf d}|$, reproduces the exact multilayer $2N \times 2N$ solutions shown in Fig.~\ref{fig1}, and it is particularly useful to discuss the features of the LL spectra (for a detailed comparison, see Methods). Without symmetry breaking, $\tilde\Delta =0$, the spectrum of $\hat H$ is valley degenerate, leading to a $4$-fold degeneracy (spin and valley) of each of the LLs in Fig.~\ref{fig1}d. In the low magnetic field range, one can also identify closely-packed groups of 3 LLs ({\em i.e.} 12-fold degenerate) whose origin we trace to three mini-valleys forming at $p \sim p_{\mathrm{c}}$ sketched in Fig.~\ref{fig1}b. At high magnetic fields, $\hat H$ generates $N$ separate groups of 4-fold degenerate low-energy LLs, which were considered to be degenerate in previous studies~\cite{min08,henni16} where the effects of $\gamma_4$ and $\delta$ were neglected, and $X(\mathbf{p}) \approx (-\kappa^{\dagger} / p_{\mathrm{c}})^N$  resulted in a Berry phase $N\pi$ singularity at $p=0$. At high fields, this group of $N$ LLs clearly separates from electron- and hole- like levels in the spectrum, whereas, at low fields, their mixing with hole-like dispersive LLs leads to an additional two-fold degeneracy for all but the $\epsilon = 0$ level.

{\bf Topological properties.}
The other notable feature of the low-energy `non-dispersive' LLs is that their states in opposite valleys are located on opposite surfaces of the film. Consequently, when asymmetry, $\Delta$, between the bottom and top surfaces is introduced by, e.g, a displacement field, $E_z$, these LLs split apart, lifting the valley degeneracy. This evolution can be traced in the LL spectrum shown in Fig.~\ref{fig1}e where $K$ and $K'$ valley states are marked using different colors. The first noticeable difference is that, now, groups of  $N$ lowest-energy LLs in opposite valleys can be traced to $\pm \Delta/2$ convergence points marked by arrows in Fig.~\ref{fig1}e. To highlight the difference in the LLs spectra in the opposite  valleys, we plot them separately in Fig.~2a, b and point out that groups of LLs that converge towards the top of the valence band (highlighted in black) slightly differ in valleys $K$ and $K'$. As these states originate from the valence band maxima, where the electron dispersion is approximately parabolic, the difference between LL spectra reflects topological properties of electron states as characterized by Berry curvature $\Omega^{(\pm)} (\mathbf{p})$ and associated magnetic moment $\mathfrak{m}_z (\mathbf{p})$~\cite{berry84,xiao10},
\begin{eqnarray}
\Omega^{(\pm)} (\mathbf{p}) &=& i \hbar^2 \langle \nabla_{\mathbf{p}} u_{\pm} | \times | \nabla_{\mathbf{p}} u_{\pm} \rangle \cdot \mathbf{\hat{e}}_z ,
= \mp \frac{\hbar^2}{2d^3} {\bf d} \cdot (\partial_{p_x} {\bf d} \times \partial_{p_y} {\bf d}), \nonumber \\
\mathfrak{m}_z (\mathbf{p}) &=& - \frac{ie\hbar}{2} \langle \nabla_{\mathbf{p}} u_{\pm} | \times [ \hat{H} - \epsilon_{\pm} ] | \nabla_{\mathbf{p}} u_{\pm} \rangle \cdot \mathbf{\hat{e}}_z \nonumber
= - \frac{e \hbar}{2d^2} {\bf d} \cdot (\partial_{p_x} {\bf d} \times \partial_{p_y} {\bf d}), \nonumber
\end{eqnarray}
where $u_{\pm}^T = {\cal N}_{\pm} \left(
                                        \begin{array}{cc}
                                          d_z \pm d , & d_x + i d_y \\
                                        \end{array}
                                      \right)$
is the sublattice Bloch spinor in the $\eps_\pm$ (conduction/valence) band, ${\cal N}_{\pm} = 1/\sqrt{2d(d \pm d_z)}$, $d = |{\bf d}|$, and $\nabla_{\mathbf{p}} = (\partial_{p_x},\partial_{p_y})$.
Note that $\Omega^{(-)} (\mathbf{p}) = - \Omega^{(+)} (\mathbf{p})$, whereas $\mathfrak{m}_z (\mathbf{p})$ is the same for the $\epsilon_{+}$ and $\epsilon_{-}$ bands.

Using the two-component model~(\ref{eq1}) for $N \gg 1$
and neglecting trigonal warping (parameters $\gamma_2$ and $\gamma_3$),
the Berry curvature $\Omega^{(\pm)}$ for conduction/valence bands is given by \cite{zhang11}
\begin{eqnarray}
\Omega^{(\pm)} (\mathbf{p}) \approx  \pm \xi \frac{2N^2 }{p^2 \gamma_1}
\frac{\tilde \Delta (\mathbf{p})  \, (p/p_{\mathrm{c}})^{2N}}{\left[ (\tilde\Delta (\mathbf{p}) / \gamma_1)^2 + 4 (p/p_{\mathrm{c}})^{2N} \right]^{3/2}} ,
\end{eqnarray}
which generalizes the result for finite $\Delta$ determined in monolayer and bilayer graphene~\cite{xiao07,xiao10}.
The orbital magnetic moment for both bands is
\begin{eqnarray}
\mathfrak{m}_z (\mathbf{p}) \approx \xi \frac{e}{\hbar}\frac{N^2}{p^2}
\frac{\tilde \Delta (\mathbf{p}) \,  (p/p_{\mathrm{c}})^{2N}}{(\tilde\Delta (\mathbf{p}) / \gamma_1)^2 + 4 (p/p_{\mathrm{c}})^{2N} } .
\end{eqnarray}
We note the strong $N^2$ layer dependence and the fact that in-plane magnetic field $\mathbf{B}_{\parallel}$ can contribute to the gap $\tilde\Delta (\mathbf{p})$.

\begin{figure}[!t]
\includegraphics[scale=0.6]{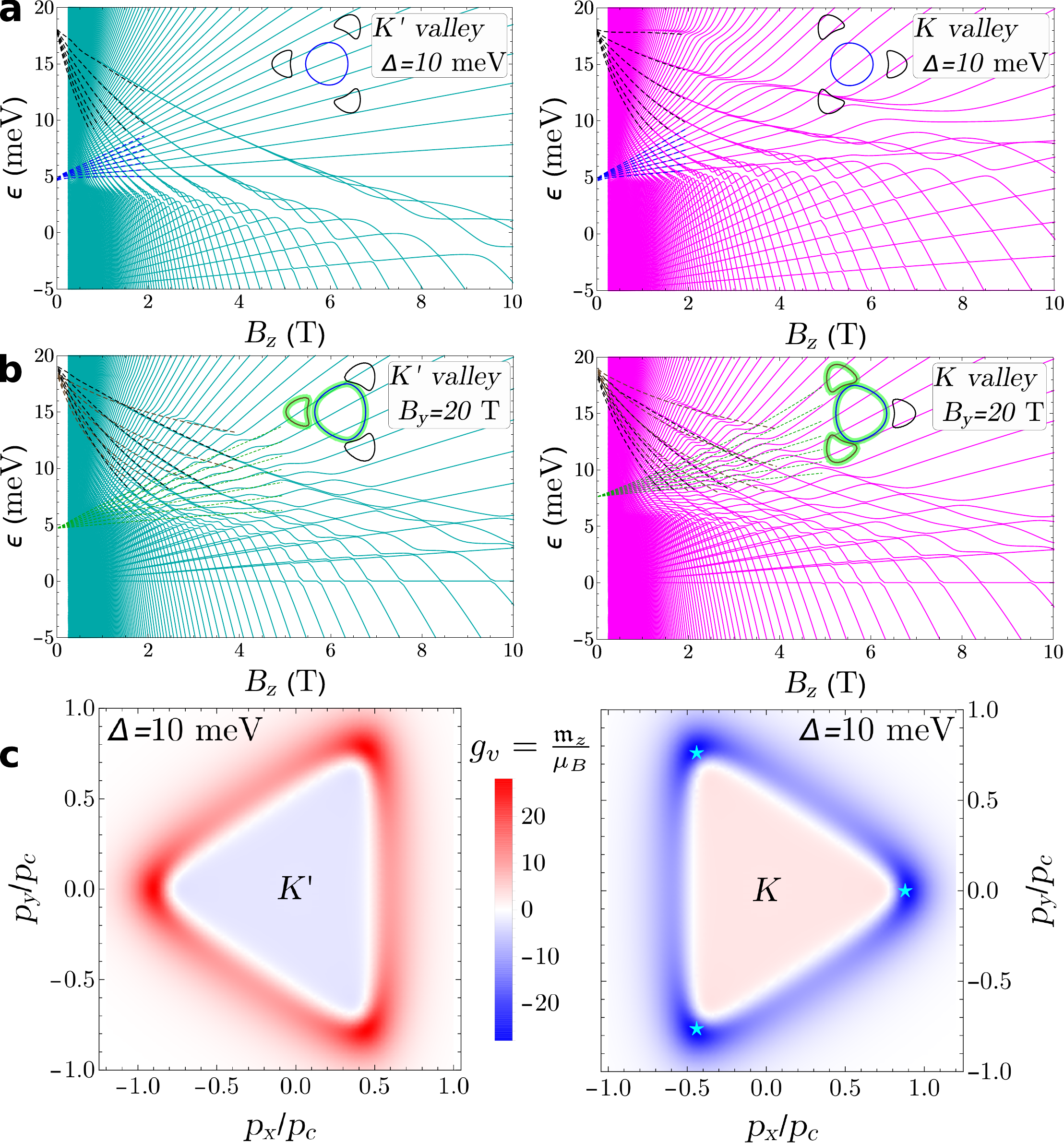}
\caption{{\bf Orbital magnetic moment and valley-dependent Landau level spectra in an ABC graphitic film with broken symmetries.} 
{\bf a} Landau level spectra as a function of perpendicular magnetic field $B_z$ in $K'$ and $K$ valleys for bottom/top asymmetry $\Delta = 10\,$meV, and for in-plane magnetic field $B_y = 20\,$T, {\bf b}. Green highlighting in the insets in {\bf b} indicate parts of electron/hole-like Fermi lines connected upon magnetic breakdown, resulting in LLs with new spacings (green dashed lines).
{\bf c}, Magnetic moment, $\mathfrak{m}_z (\mathbf{p})$, of Bloch states in valleys $K'$ and $K$ for $\Delta = 10\,$meV, showing opposite sign and orientation in the two valleys and magnitude peaked near the valence band mini-valleys at $p \approx p_{\mathrm{c}}$.
}\label{fig2}
\end{figure}

Whereas $\Omega^{(\pm)} (\mathbf{p})$ and $\mathfrak{m}_z (\mathbf{p})$ are peaked at $p \approx 0$ in monolayer and bilayer, for rhombohedral graphene with $N \gg 1$ they are peaked at $p \approx p_{\mathrm{c}}$. Moreover, trigonal warping (parameters $\gamma_2$ and $\gamma_3$) introduces anisotropy such that $\Omega^{(\pm)} (\mathbf{p})$ and $\mathfrak{m}_z (\mathbf{p})$ are peaked in the vicinity of the valence band maxima.
This is demonstrated in Fig.~2c where we plot the distribution of magnetic moment $\mathfrak{m}_z$ across momentum space in both valleys for $\Delta = 10\,$meV. This  magnetic moment leads to a valley splitting, $2 \mathfrak{m}_z B_z$, ({\em e.g.} of the LL converging towards the valence band edge, Fig.~2a), that can be characterized by a factor $g_{\mathrm{v}} = \mathfrak{m}_z/\mu_{\mathrm{B}}$ which can be
as large as $g_{\mathrm{v}} \sim 100$, Fig.~3a.  An application of in-plane magnetic field would additionally lift the degeneracy between the three valence band maxima, as illustrated using the LL fan in Fig.~2b.

\begin{figure}[!t]
\includegraphics[scale=0.55]{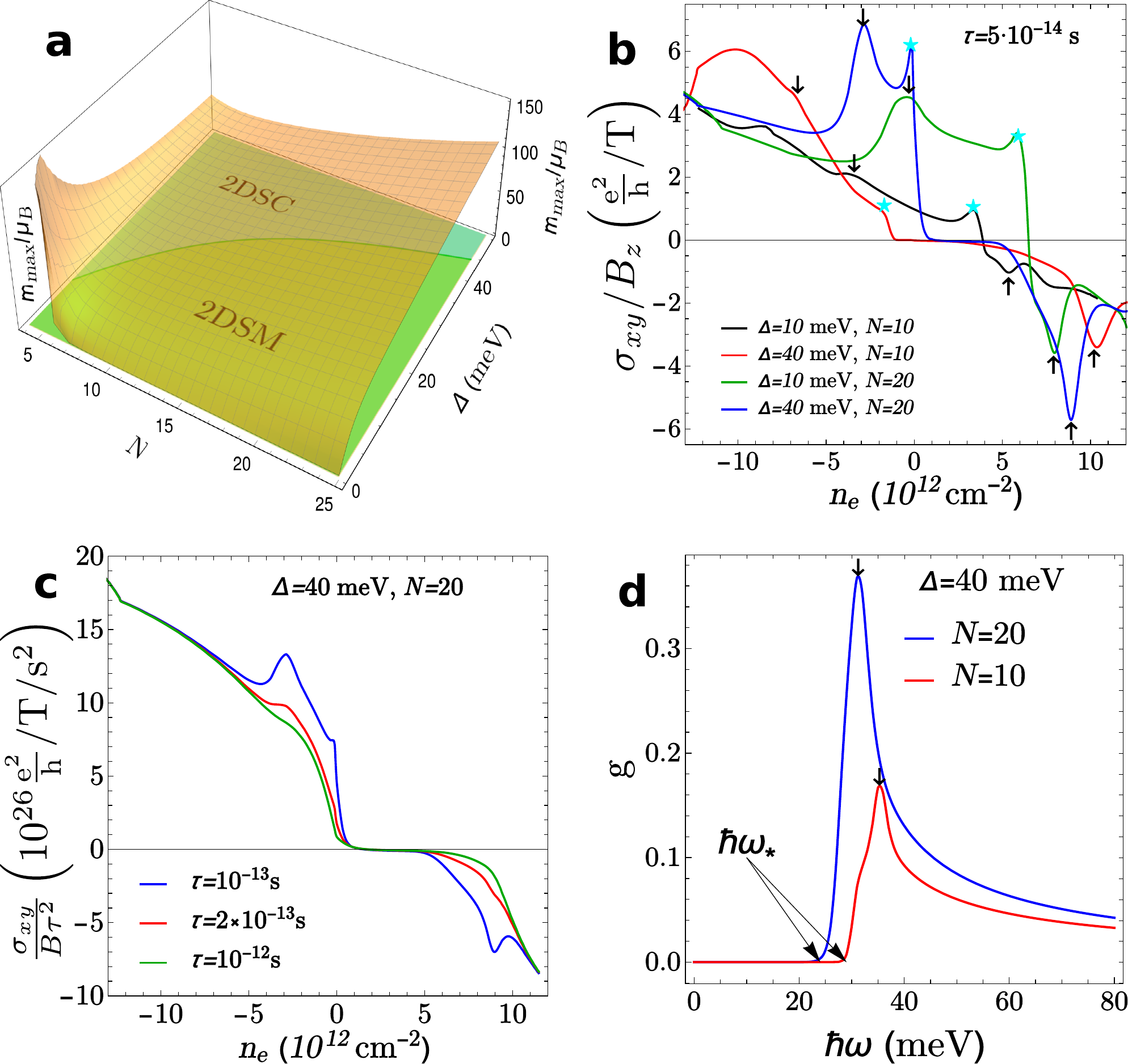}
\caption{{\bf Anomalous Hall effect.} {\bf a}, Maximum magnitude of the magnetic moment $\mathfrak{m}_z$ in valley $K$ in units of Bohr magneton $\mu_{\mathrm{B}}$, plotted as a function of the number of layers, $N$, and asymmetry $\Delta$; the basal plane identifies the parametric range where the top/bottom layer asymmetry transforms a 2DSM into a semiconductor (2DSC).
{\bf b}, Total Hall coefficient $\sigma_{xy} = \sigma_{xy}^{\mathrm{A}} + \sigma_{xy}^{\mathrm{H}}$ per unit magnetic field $B_z$ for scattering time $\tau = 5\times10^{-14}\,$s plotted as a function of carrier density $n_{\mathrm{e}}$ where $\sigma_{xy}^{\mathrm{A}}$ is the anomalous Hall coefficient and $\sigma_{xy}^{\mathrm{H}}$ is the classical contribution (parameters according to the legend).
{\bf c}, Total Hall coefficient $\sigma_{xy}$ per unit magnetic field $B_z$ for $\Delta = 40 \, \rm meV$, $N=20$, plotted for different choices of scattering time and normalized by $\tau^2$.
{\bf d}, Absorption coefficient $g (\omega)$ for an ABC graphitic film in the 2DSC range, as a function of photon energy $\hbar\omega$, calculated in the full multi-band model (see discussion in Methods).
}\label{fig3}
\end{figure}

{\bf Anomalous Hall effect.}
In the small magnetic field range where  electron dynamics can be described classically (rather than using Landau quantization), the splitting of magnetic moments  at the valence band edges in the opposite valleys would shrink the size of the Fermi pockets for the holes in one valley and expand them in the other. The resulting valley polarization of holes would manifest itself in transport characteristics. This is because,
in the presence of in-plane electric field~${\bf E}$, carriers in bands with Berry curvature experience a  drift ~\cite{xiao10,nagaosa10},
\begin{eqnarray*}
{\bf v}_{\pm}(\mathbf{p}) = \nabla_{\mathbf{p}} \epsilon_{\pm} (\mathbf{p}) + (e/\hbar){\bf E} \times \mathbf{\hat{e}}_z \Omega^{(\pm)} (\mathbf{p}) , \label{vel}
\end{eqnarray*}
so that, together, Berry curvature and valley polarization produce an anomalous contribution to the Hall effect,
\begin{eqnarray}
\sigma_{xy}^{\mathrm{A}} = -\frac{e^2B_z}{\pi^2\hbar^3} \sum_{i} \oint_{{\cal L}_{i} (K)} \frac{\Omega_{i}^{(\pm)} \mathfrak{m}_{i,z} } {|\nabla_{\mathbf{p}} \epsilon_{i}|} dp . \label{ah}
\end{eqnarray}
Here $i$ lists all Fermi lines ${\cal L}_{i}$ in valley $K$ for both $\epsilon_{\pm}$ bands, and the linear integral is taken in the anticlockwise direction along each Fermi line, ${\cal L}_{i}$.
When combined with the classical kinetic Hall coefficient,
\begin{eqnarray*}
\sigma_{xy}^{\mathrm{H}} = -\frac{e^3\tau^2}{\pi^2 \hbar^2} \sum_{i,\gamma} \oint_{{\cal L}_{i} (K)} \frac{(\partial_{p_x}\epsilon_{i})}{|\nabla_{\mathbf{p}} \epsilon_{i}|}   [\nabla_{\mathbf{p}} \epsilon_{i} \times {\bf B}]_{\gamma} \frac{dp}{m_{\gamma y}} ,
\end{eqnarray*}
computed for the same band structure ($m_{\alpha \beta}^{-1} = \tfrac{\partial^2 \epsilon_{i}}{\partial p_{\alpha}\partial p_{\beta}}$) and elastic scattering time $\tau = 5 \times 10^{-14}\,$s, this yields the overall Hall conductivity, $\sigma_{xy}$, displayed in Fig.~3b.
The anomalous contribution is most pronounced for small scattering times, $\tau \sim 10^{-13}\,$s, otherwise the classical kinetic Hall conductivity dominates, as shown in Fig.~\ref{fig3}c, where we plot the ratio $\sigma_{xy}/(B \tau^2)$ for $\tau = 10^{-13},  2 \times 10^{-13}, 10^{-12}\, \rm  s$  [note that the value obtained for $\tau=10^{-12} \, \rm s$ is indistinguishable from $\sigma^{\mathrm{H}}_{xy}/(B \tau^2) $].

The doping density dependence of the  overall Hall conductivity, shown in Fig.~3b for ten- and twenty-layer films with $\Delta = 10$ meV and $40$ meV, carries certain features reflecting the distribution of Berry curvature and magnetic moment (Fig.~2c) across the electron dispersion. According to the parametric plot shown on the basal plane of Fig.~3a, both ten- and twenty-layer films with $\Delta = 40$ meV are gapful 2D semiconductors (2DSC), so that the states with the maximium $\mathfrak{m}_z\Omega$ near the valence band edge (Fig.~1c) are reached at a relatively small p-doping, resulting in a hump in $\sigma_{xy}(n_{\mathrm{e}})$ at $n_{\mathrm{e}}<0$ indicated by a star. For a smaller gap, the films appear to be 2D semimetals (2DSM), so that the part of the band where $\mathfrak{m}_z\Omega$  is the  largest is set above the Fermi level for undoped materials, shifting the hump in $\sigma_{xy}$  to  $n_{\mathrm{e}}>0$.   A double-peak structure is caused by the convolution of the ${\mathfrak m}_z\Omega$ product and the density of states with a pronounced Van Hove singularity (pointed by down arrows) in Eq.(\ref{ah}). At higher electron dopings,  the Fermi level reaches the maximum of ${\mathfrak m}_z\Omega$ for the conduction band, causing a negative peak in Hall coefficient (pointed by up arrows). 

Pumping inter-band transitions with circularly polarized photons induces
a partial valley polarization in graphene and graphite (thus breaking the time-inversion symmetry).
Combined with the Berry curvature effect, this would produce a Hall-like drift current (at $B_z = 0$) perpendicular to the static in-plane electric field, which can be characterized by an anomalous transverse photoconductivity, $\delta\sigma_{xy}$, which appears to be especially pronounced for ABC graphitic films in the 2DSC regime.
This effect is determined by the resonant inter-band absorption of circularly-polarized photons with energy $\omega$ and absorption coefficient $g(\omega)$, plotted in Fig.~3d,
\begin{eqnarray}
\delta \sigma_{xy}  &=& -\zeta \frac{e^2}{\hbar} \frac{W}{ \hbar \omega} \tau_{\mathrm{rec}} |\Omega^-_{K}(p_{\mathrm{max}})| g (\omega) ,
\end{eqnarray}
where $\zeta=\pm 1$ stands for left/right handed circular polarization of the pump (radiation is approaching from the top) with power density $W$, $\Omega^-_{K}$ is the Berry curvature of the valence band at valley $K$, and $\tau_{\mathrm{rec}}$ is the lifetime of photo-excited holes at the top, $p_{\mathrm{max}}$, of the valence band in the photo-activated valley.

\vspace{1.0cm}
{\bf Discussion}

\begin{figure}[!t]
\includegraphics[scale=0.7]{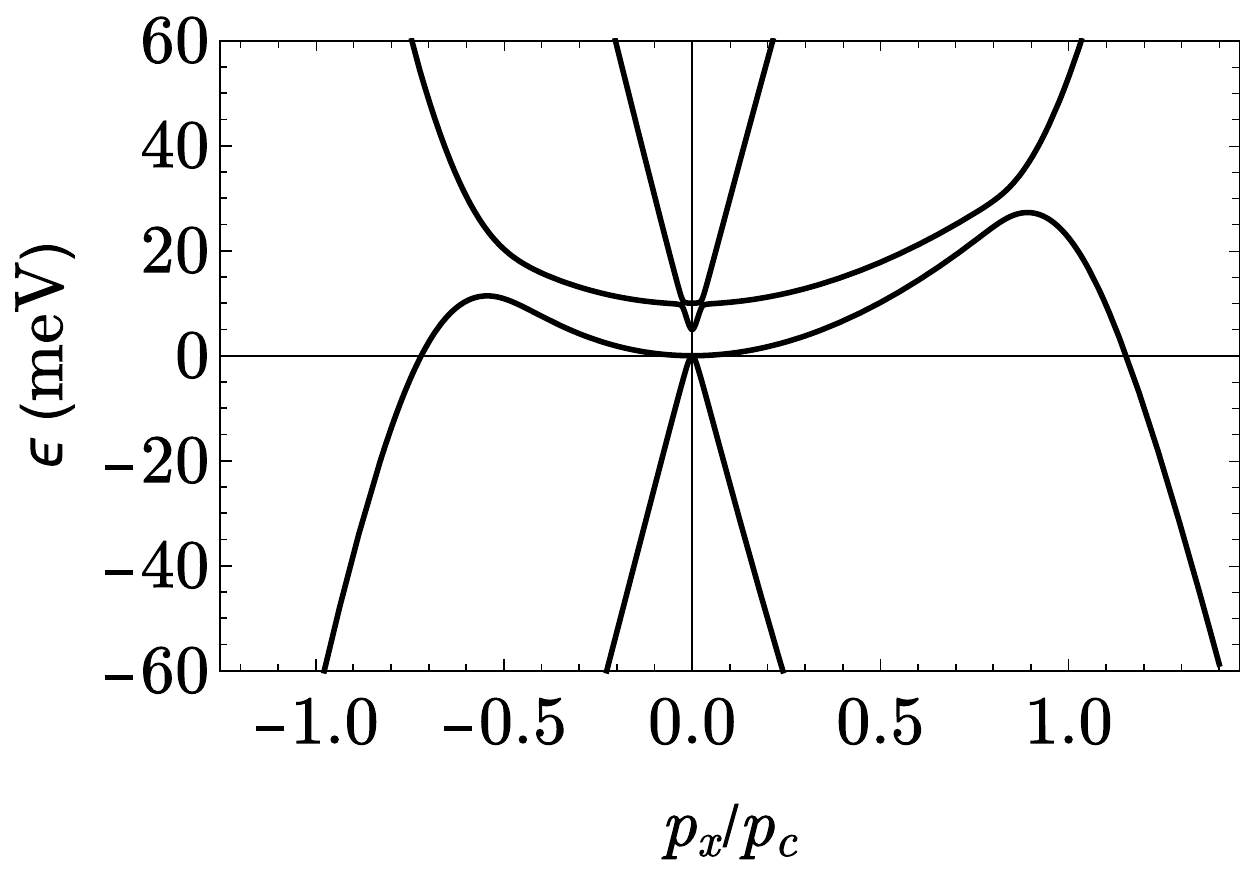}
\caption{{\bf Rhombohedral graphite film with a surface stacking fault.}  Low-energy dispersion $\epsilon (p_x,0)$ in the $K$ valley of a rhombohedral graphite film with $N = 11$ layers and an ABA stacking fault at the surface. Here we use the same parameters as in Fig.~1 with $\Delta = 0$ and $\delta = 10 \, \rm meV$ (see Methods for details).
}\label{fig:Defect}
\end{figure}

We have shown that thin films of rhombohedral graphite with up to tens of layers of thickness host two-dimensional electron states characterized by a large Berry curvature and a giant intrinsic magnetic moment. Note that stacking faults~ \cite{garciaruiz19} in an rhombohedral film give rise to an additional strongly dispersing band that simply overlays the surface states spectrum. For example, in Fig.~\ref{fig:Defect} we show the spectrum of an eleven layer rhombohedral (ABC) film with an ABA stacking fault at the top surface, which features a graphene-like Dirac band with velocity $v/\sqrt{2}$ and a small asymmetry gap, determined by~$\delta$.

Semiconducting ABC graphitic films may be also used to create topologically-protected edge modes at interface regions with an inverted sign of $\Delta$, controlled,{\it e.g.}, by an oppositely-directed displacement field. Similarly to a $\pm \Delta$ domain wall in bilayer graphene~\cite{martin08},  a $\pm \Delta$ interface in $N$ layer ABC graphite would host $N$ co-propagating one-dimensional bands inside the spectral gap (with opposite direction of propagation  in $K$ and $K'$ valleys).
Also, an atomic step in the film thickness may produce an isolated pair of edge states inside the gap (one in each valley with opposite directions of drift), but this feature will depend on the crystallographic orientation of the edge: it would be best developed for a zigzag termination of the additional layer, and  it would be suppressed for the armchair edge due to valley mixing.
Finally, as the gapful spectrum of an ABC film may be the result of many body effects leading to spontaneous spin/valley symmetry breaking~\cite{lemonik10,jung11}, topological features of ABC graphite, enhanced by the $N \gg 1$ number of layers, would produce various possibilities for gapless edge modes in the system.
For valley-symmetric magnetic phases, this would result in valley-current carrying domain walls, hosting $N$ one-dimensional channels with the opposite direction of drift in valleys $K$ and $K'$.
For phases with a valley antisymmetric order parameter, this would result in $2N$ electrical-current carrying edge modes at the physical edge of the sample.
Overall, multilayer rhombohedral graphite films offer an attractive playground for studying giant topological effects in their electronic transport characteristics.

\begin{figure}[!t]
\includegraphics[scale=0.19]{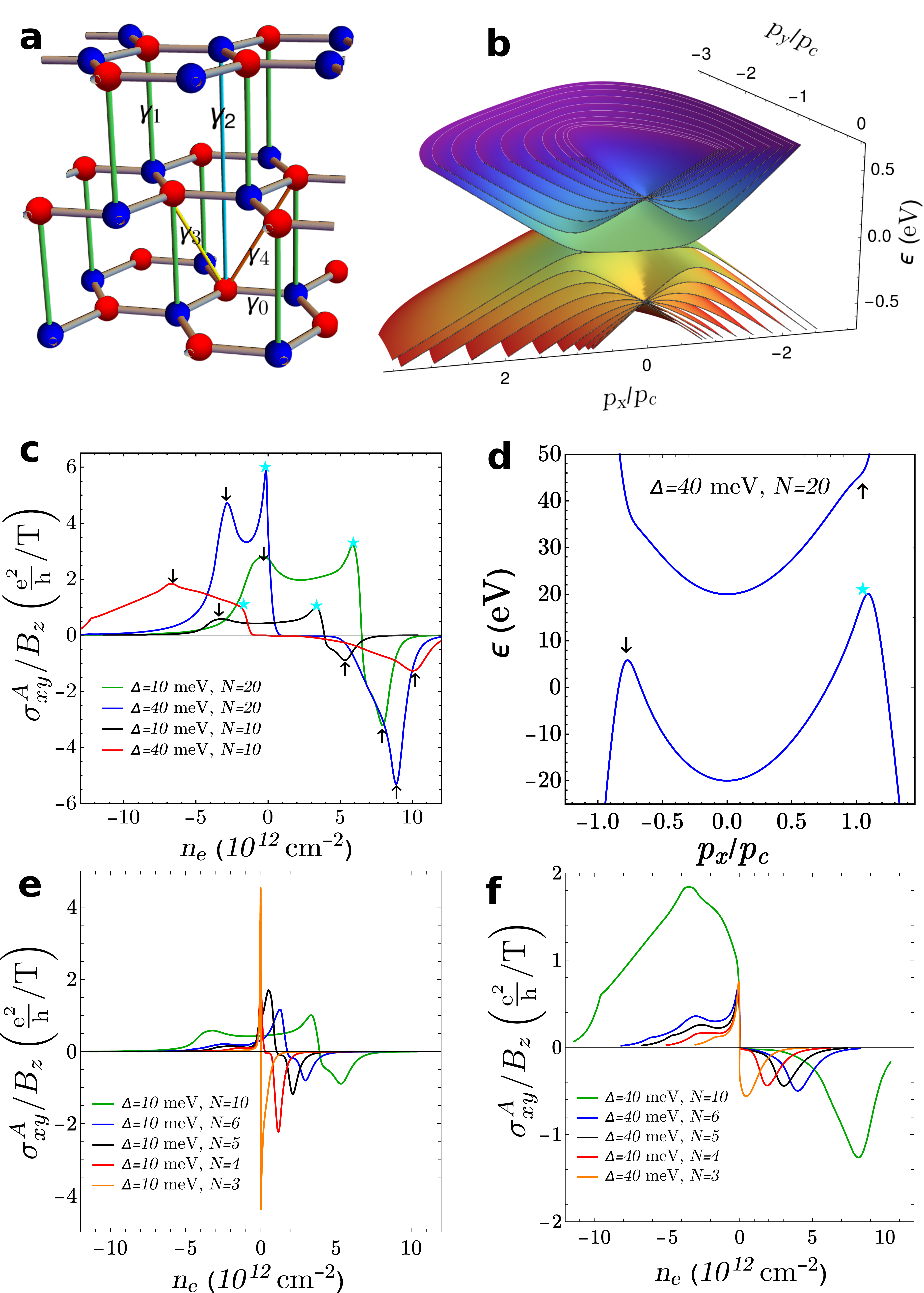}
\caption{{\bf Rhombohedral graphite film spectrum with the full band model and anomalous Hall coefficient:} {\bf a}, Hopping parameters in the Slonzewski-Weiss-McClure model;
{\bf b}, subband spectrum $\epsilon (p_x,p_y)$ of 10-layer ABC graphite; {\bf c} Anomalous Hall contribution  $\sigma_{xy}^{\mathrm{A}}$ to the Hall effect per unit magnetic field $B_z$ plotted as a function of carrier density $n_{\mathrm{e}}$. Black down arrows indicate van Hove singularities,  cyan stars and black up arrows indicate the peaks originating  from hot spots of Berry curvature in the valence/conduction bands, as shown in the dispersion plot {\bf d}. {\bf e, f} anomalous Hall contribution $\sigma_{xy}^{\mathrm{A}}$ per unit magnetic field $B_z$ for a smaller number $N$ of layers. As expected from Fig.~3a,  the anomalous effect is maximal at low $\Delta$ for $N\lesssim 6$,  while it grows with $\Delta$ for $N\gtrsim 6$.
}\label{Extfig1}
\end{figure}

\vspace{1.0cm}
{\bf Methods}

{\bf Hybrid $\mathbf{k} \cdot \mathbf{p}$ theory and tight-binding model.}
Our calculations are based on the hybrid $\mathbf{k} \cdot \mathbf{p}$ theory and tight-binding model (HkpTB). HkpTB combines the expansion of the electronic Hamiltonian for a single graphene layer in the lowest relevant order of momentum counted from the center of the $K$ and $K^{\prime}$ valleys, which we use in the form $\kappa = \xi p_x + i p_y + i e x B_z/\hbar$, with interlayer hopping that takes into account the lattice arrangement for rhombohedral graphite displayed in Fig.~\ref{Extfig1}a. Here we use a $2N$ component basis $(\psi_{\mathrm{A1}}, \psi_{\mathrm{B1}}, \psi_{\mathrm{A2}}, \psi_{\mathrm{B2}}, \cdots, \psi_{\mathrm{AN}}, \psi_{\mathrm{BN}})$ of $A$ and $B$ sublattice Bloch states for each valley and take into account all hoppings in the Slonczewski-Weiss-McClure parametrization of graphite~\cite{mcclure69,dressel02,koshino09}, as marked in Fig.~\ref{Extfig1}a. The Hamiltonian that determines the electronic bands in $N$-layer rhombohedral graphite reads
\begin{eqnarray}
 \hat H_{N} =
\begin{pmatrix}
 D_1 & V_{1,2} & W & 0 & \cdots \\
 V_{1,2}^{\dagger} & D_2 & V_{2,3} & W & \cdots \\
W^\dagger & V_{2,3}^\dagger & D_3 & V_{3,4} & \cdots \\
0 & W^\dagger & V_{3,4}^\dagger & D_4 & \cdots  \\
\vdots & \vdots & \vdots & \vdots & \ddots
\end{pmatrix},
\label{HNfull}
\end{eqnarray}
where we use $2 \times 2$ blocks
\begin{eqnarray*}
D_1 \!\!&=&\!\!
\begin{pmatrix}
 U_1 & v \kappa_{1,1}^\dagger
 \\ v \kappa_{1,1} & U_1 + \delta
\end{pmatrix} ,
\quad D_N =
\begin{pmatrix}
U_N + \delta & v \kappa_{N,N}^\dagger
 \\ v \kappa_{N,N} & U_N
\end{pmatrix} ,
\\
D_n \!\!&=&\!\!
\begin{pmatrix}
 U_n + \delta & v \kappa_{n,n}^\dagger
 \\ v \kappa_{n,n} & U_n + \delta
\end{pmatrix} \quad (n=2,3, \ldots , N-1), \label{di}
\\
V_{n,n+1} \!\!&=&\!\!
\begin{pmatrix}
-v_4 \kappa_{n,n+1}^\dagger & v_3\kappa_{n,n+1} \\ \gamma_1 & -v_4 \kappa_{n,n+1}^\dagger
\end{pmatrix},
\quad
W =
\begin{pmatrix}
0 & \tfrac{1}{2}\gamma_2 \\ 0 & 0
\end{pmatrix} \\
\kappa_{n,n} & =& \kappa  + \tfrac{1}{2}e a_z (2n - N - 1)(\xi B_y - i B_x) \, ,\\
\kappa_{n,n+1} &=& \kappa + \tfrac{1}{2}e a_z (2n - N)(\xi B_y - i B_x) \, .
\end{eqnarray*}
Here $a_z$ is the interlayer spacing, $v = (\sqrt{3}/2)a\gamma_0/\hbar$, $v_3 = (\sqrt{3}/2)a\gamma_3/\hbar$, $v_4 = (\sqrt{3}/2)a\gamma_4/\hbar$ and $a$ is the in-plane lattice constant.
The intra ($D_n$) and interlayer ($V_{n,n+1}$) elements take into account the in-plane components $B_x$, $B_y$ of an arbitrarily-oriented magnetic field via the Peierls substitution, generalising an approach applied previously to bilayer graphene~\cite{kheirabadi16}.
Matrix element $W$ describes next-neighboring-layer hopping in the vertical direction and $\delta$ represents a difference in energy between the dimer sites inside the crystal and non-dimer sites $A_1$ and $B_N$ in the outer layers.

The minimal tight-binding model, consisting of only nearest-neighbour intra- and interlayer hopping $\gamma_0$ and $\gamma_1$, approximates $X ({\bf p}) \approx (-\kappa^{\dagger}/p_{\mathrm{c}})^N$ in Eq.~(\ref{eq1}) yielding flat, isotropic surface bands $\epsilon_{\pm} \approx (p/p_{\mathrm{c}})^N$.
Skew interlayer hopping $\gamma_3$ and next-nearest-layer hopping $\gamma_2$ produce trigonal warping as shown in the dispersion Fig.~\ref{fig1}a and the Fermi surface plots in Fig.~\ref{fig1}b. Skew interlayer hopping $\gamma_4$ and parameter $\delta$ break electron-hole symmetry in the energy spectrum.
For numerical diagonalization of Eq.~\ref{HNfull} we use the following values of tight-binding parameters~\cite{kuz09} $\gamma_0 = 3.16\,$eV, $\gamma_1 = 0.381\,$eV, $\gamma_3 = 0.38\,$eV, $\gamma_4 = 0.14\,$eV, $\gamma_2 = - 0.020\,$eV~\cite{dressel02} and $\delta = 0.0\,$eV.

The band structure near the valley center is plotted in Fig.~\ref{Extfig1}b, obtained by diagonalization of the Hamiltonian in Eq.~\ref{HNfull} for $N = 10$ layers. There are $N-1$ conduction/valence bands with apex at $p=0$ and energy $\pm \gamma_1$ arising from the interlayer hybridization of carbon orbitals within the middle of the graphite film. Additionally, there are two flat bands at the conduction/valence band edge with wave functions made of $p_z$ carbon orbitals on sites $A_1$ and $B_N$ in the surface layers (the ones that don't find a nearest neighbor in the next layer to hybridize into a dimer state). Zooming into the dispersion of the flat bands at low energy reveals semi-metallic behavior in Fig.~\ref{fig1}a and parametric regimes identified in Fig.~\ref{fig1}b. The full band model was used to compute the anomalous Hall coefficient using the formal definitions of $\Omega^{(\pm)} (\mathbf{p})$ and $\mathfrak{m}_z (\mathbf{p})$ with the results in Fig.~\ref{Extfig1}c; Fig.~\ref{Extfig1}d indicates how features in the low-energy dispersion are related to peaks in the anomalous Hall effect. Fig.~\ref{Extfig1}e, f shows the anomalous Hall effect for a smaller number of layers with different asymmetry values.

\begin{figure}[!t]
\includegraphics[scale=0.59]{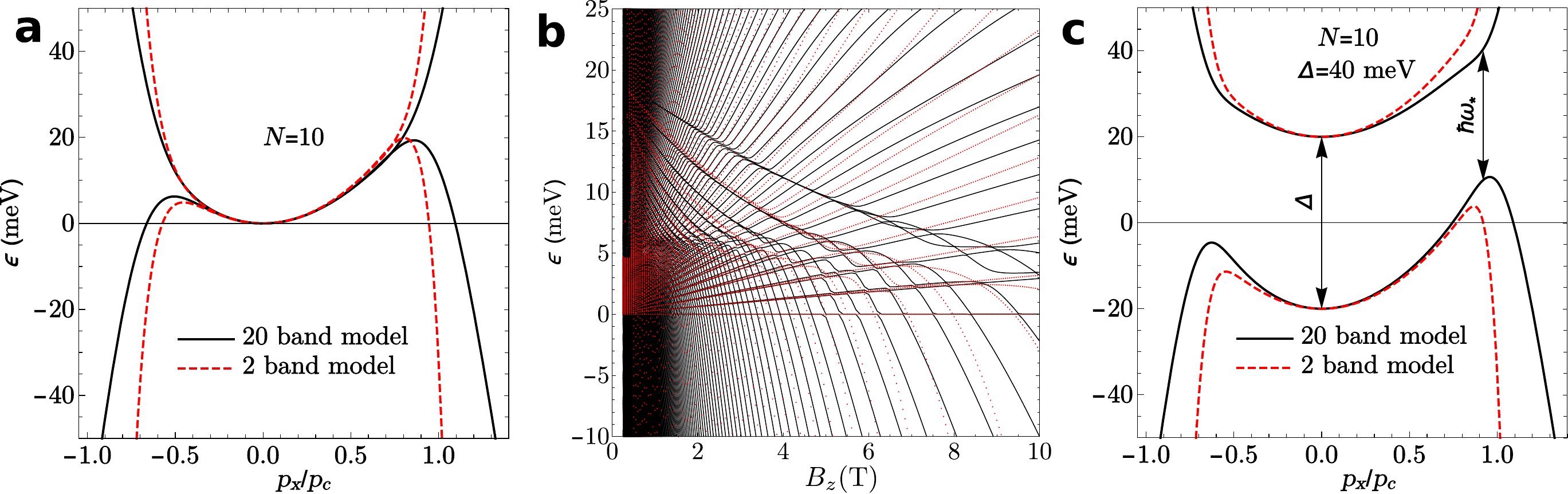}
\caption{{\bf Comparison of the full band model with the effective two-band model:} {\bf a}, Low-energy dispersion $\epsilon (p_x,0)$ for $B=0$ and $\Delta=0$.
{\bf b}, Landau level spectrum as a function of perpendicular magnetic field $B_z$ for $\Delta=0$. {\bf c}, Low-energy dispersion $\epsilon (p_x,0)$ for  $B=0$ and $\Delta=40\, \rm meV$. In {\bf c}, we indicate the threshold $\hbar \omega_*$ for photo-absorption which appears to be lower than $\Delta$ when calculated in the multi-band model (similarly to bilayer graphene~\cite{mcc06}).
}\label{Extfig2}
\end{figure}

The simplified $2 \times 2$ (two band) Hamiltonian in Eq.~1 was derived using a Schrieffer-Wolff transformation which eliminated all high energy conduction and valence subbands projecting Hamiltonian $\hat H_{N}$ onto a basis of $(\psi_{\mathrm{A1}} , \psi_{\mathrm{BN}})$ Bloch states using perturbation theory in intra- and interlayer hoppings up to the $N$th order with effective mass $m_{\star}^{-1} = (2v^2 / \gamma_1) (2\gamma_4/\gamma_0 + \delta/\gamma_1) \sim (0.4 m_0)^{-1}$ where $m_0$ is the free electron mass.
The energy spectrum predicted by the two band model Eq.~\ref{eq1} is compared with that of the full HkpTB Eq.~\ref{HNfull} in Fig.~\ref{Extfig2}a for zero magnetic field and zero asymmetry, in Fig.~\ref{Extfig2}b as a function of perpendicular magnetic field $B_z$ with zero asymmetry, and in Fig.~\ref{Extfig2}c for zero magnetic field and finite asymmetry $\Delta=40\, \rm meV$.

To study the Landau level spectra in a perpendicular magnetic field ${\bf B} = (0,0,B_z)$ we use the Landau gauge for the vector potential ${\bf A} = (0,B_z x,0)$.
Then, $\kappa$ and $\kappa^{\dagger}$ transform into
raising and lowering operators for magnetic oscillator states $\phi_n$ in valley $K$~\cite{mcc06,koshino09} (vice versa in valley $K^{\prime}$) with $\kappa \phi_n = i (\hbar / \lambda_B) \sqrt{2(n+1)} \phi_{n+1}$ and $\kappa^{\dagger} \phi_n = -i (\hbar / \lambda_B) \sqrt{2n} \phi_{n-1}$ with $\lambda_B = \sqrt{\hbar / (eB_z)}$.
Numerical diagonalization of the Hamiltonian in Eq.~\ref{HNfull} is performed in a basis with a series of oscillator states for each sublattice component, $(\phi_{0} , \phi_{1} , \cdots , \phi_{N_0 + n - 1})$ for $A_n$ and $(\phi_{0} , \phi_{1} , \cdots , \phi_{N_0 + n})$ for $B_n$, where $N_0 = 250$ is a cutoff, sufficient for convergence for $B>0.5\,$T. The result of the exact diagonalization is shown in Fig.~1d and e, indicating that the two-band model is sufficient to catch the qualitative features in the behavior of LLs at small magnetic fields, but with substantial quantitative deviations developing at $B_z > 8\,$T.

{\bf Estimate of the bulk band gap.}
For an infinite number of layers (3D rhombohedral graphite), the bulk bands are gapless
at $p=p_{\mathrm{c}}$ (the position of the Dirac point follows a continuous spiral as a function of the out-of-plane wave vector~\cite{heik11,ho13,ho14,ho16,hyart18}).
In order to estimate the magnitude of the bulk gap for finite $N$, we
consider the minimal model (including only $\gamma_0$ and $\gamma_1$) with in-plane momentum of magnitude $p = p_{\mathrm{c}}$
and directed in the $x$ direction only, {\em i.e.} $\mathbf{p} = (p_{\mathrm{c}}, 0)$.
Then, the $2N \times 2N$ Hamiltonian [Eq.~\ref{HNfull}
with $\gamma_2 = \gamma_3 = \gamma_4 = \delta = U_i = |\mathbf{B}|=0$] may be written
as
\begin{eqnarray*}
 \hat H_{N} = \gamma_1
\begin{pmatrix}
0 & 1 & 0 & 0 & 0 &\cdots \\
1 & 0 & 1 & 0 & 0 &\cdots \\
0 & 1 & 0 & 1 & 0 &\cdots \\
0 & 0 & 1 & 0 & 1 &\cdots  \\
0 & 0 & 0 & 1 & 0 &\cdots  \\
\vdots & \vdots & \vdots & \vdots & \vdots & \ddots
\end{pmatrix},
\end{eqnarray*}
This is simply the tight-binding model of a linear chain with
$2N$ sites and nearest-neighbor hopping $\gamma_1$ between every site.
The solutions are $E = 2 \gamma_1 \cos ( \pi j / (2N+1))$
for $j = 1 , 2 , \ldots , 2N$ which describes an electron-hole symmetric
series of energies. The positive energy closest to zero ($j = N$) which
describes the surface state
is $E_0 = 2 \gamma_1 \cos (\pi N / (2N+1)) = 2 \gamma_1 \sin (\pi  / [2(2N+1)])$
and $E_0 \approx \pi \gamma_1 / (2N)$ for $N \gg 1$.
The positive energy which is the next closest to zero ($j = N-1$)
which represents the lowest-energy bulk band
is $E_1 = 2 \gamma_1 \cos (\pi (N-1) / (2N+1)) = 2 \gamma_1 \sin (3\pi  / [2(2N+1)])$
and $E_1 \approx 3\pi \gamma_1 / (2N)$ for $N \gg 1$.
Thus, we estimate the bulk band gap $\Delta_{\mathrm{bulk}} = 2E_1$ to be
\begin{eqnarray}
\Delta_{\mathrm{bulk}}  \approx 3\pi \gamma_1 / N  \label{GapEstimate},
\end{eqnarray}
for $N \gg 1$.
We estimate $\Delta_{\mathrm{bulk}} \approx 120\,$meV
for $N=30$ which seems to be in agreement with the numerical calculation of
Henni et al~\cite{henni16}, and we estimate $\Delta_{\mathrm{bulk}} \approx 36\,$meV for $N = 100$.

{\bf Semiclassical quantization in a magnetic field and magnetic breakdown.}
For a given cyclotron orbit $C$ with area $S(C)$ in reciprocal space~\cite{xiao10}, the semiclassical quantization condition is
\begin{eqnarray}
S(C) = \frac{2\pi}{\lambda_B^2} \left( n + \frac{1}{2} - \frac{\Gamma(C)}{2\pi} \right) ,
\label{quantization}
\end{eqnarray}
and $\Gamma(C)$ is the Berry phase of orbit $C$ defined as the
$E = \eps(\mathbf{p}) - \mathbf{B} \cdot \mathfrak{m} (\mathbf{p})$ contour.
We find that for the $K$ valley and $\Delta>0$, the Landau levels are described by \eq{quantization} with $n = 0,1,2...$, while for $\Delta<0$ the levels are described by \eq{quantization} with $n = 1,2...$, see Fig.~\ref{fig2}a.
Similarly, taking into account that parallel magnetic field creates an effective top-bottom asymmetry gap $\sim [ \mathbf{p} \times \mathbf{B}_{\parallel} ]_z$  (see \eq{eq1}), for $B_y = 20\,$T, Landau levels of the $p_x>0$ conduction band pocket are described in the K valley by \eq{quantization} with $n = 1,2...$, while the two pockets with $p_x<0$ are described by \eq{quantization} with $n = 0,1,2,..$. These rules are swapped in the other valley (or, for $B_y = -B_y$ ), see Fig.~\ref{fig2}b.

\begin{figure}[t]
\includegraphics[scale=0.4]{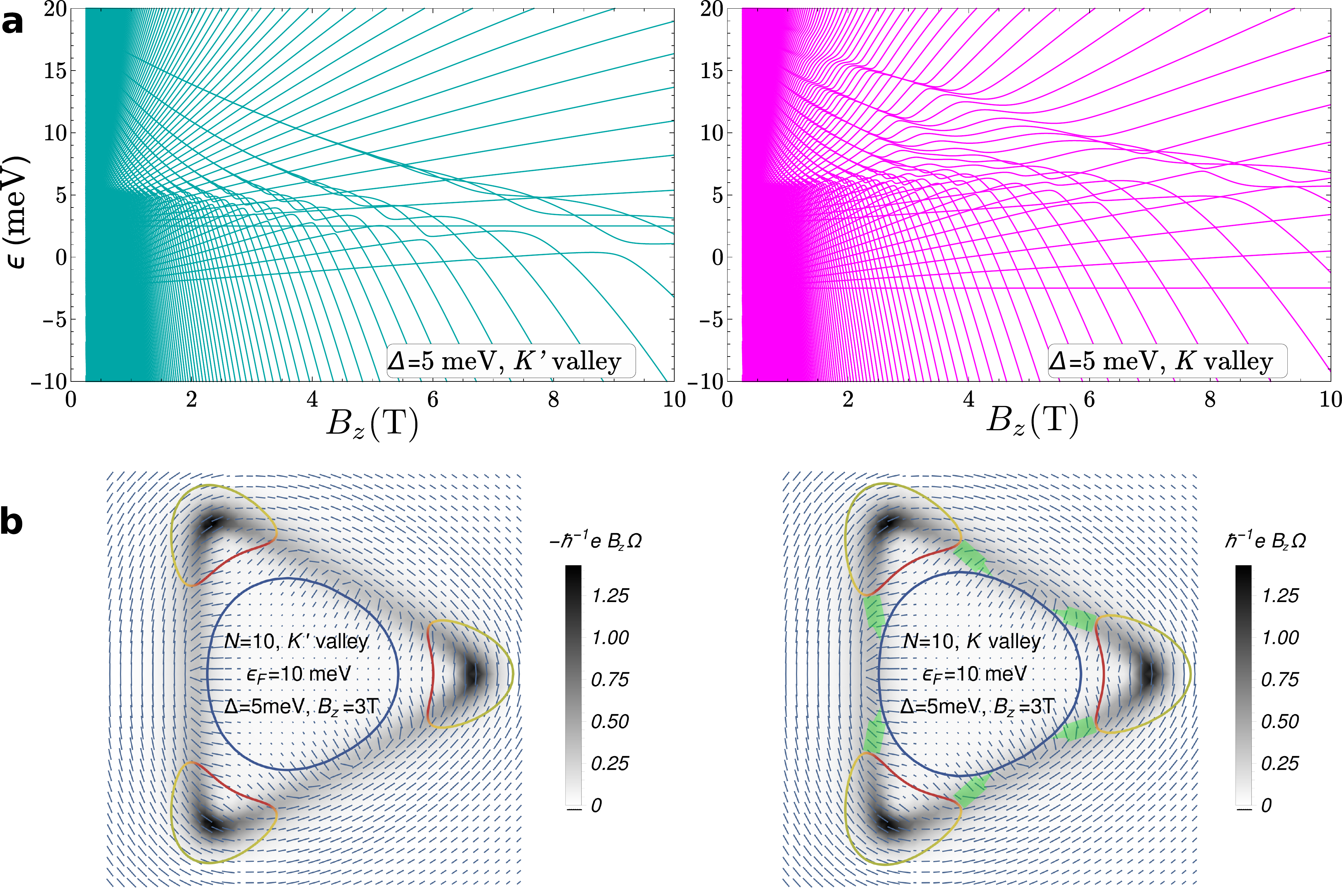}
\caption{{\bf a}, Landau level spectra as a function of perpendicular magnetic field $B_z$ for $\Delta = 5\, \rm meV$ showing strong interband magnetic  breakdown for the $K$ valley  and almost none for the $K'$ valley. {\bf b}, Fermi surfaces offset by orbital magnetic moment at $B_z = 3 $T are plotted for $\Delta = 5\,$meV at valley $K'$/$K$. Red/blue color of the contours indicates the amplitude of the wave-function on the top/bottom layers, the gray level background shows the Berry curvature for the valence band, that is positive in the $K$ valley and negative in the $K'$ valley.  The blue vector field lines correspond to the value and direction of the maximal gradient $\<u_{\mathrm{v}}(p)| \nabla_p |u_{\mathrm{c}}(p)\>$. Notable magnetic breakdown (indicated by thick green arrows) happens only in the $K$ valley, where the factor $1 + e \Omega  B_z/\hbar$ increases the breakdown phase space.
\label{fig:Explanation}
}
\end{figure}

As is evident from Fig.~\ref{fig2}b for $\Delta=10\, \rm meV$  and  Fig.~\ref{fig:Explanation}a for $\Delta=5 \,\rm meV$, there is a significant mixing of LLs of the valence and conduction bands for valley $K$,  while such mixing is almost absent in valley $K'$ (this is reversed for $\Delta\to -\Delta$).
At zero magnetic field, the states in the valence and conduction bands are orthogonal for a given quasimomentum, while a semiclassical wave-packet drifting along the Fermi contour at non-zero magnetic field has non-zero spread of quasimomenta around the Fermi contour, leading to non-zero projection onto another band. Thus,  there are two main factors that contribute to interband breakdown: (i) the extent to which the states in the two bands are non-orthogonal when taken at nearby points in momentum space, and (ii) the time that the wave-packet spends in the part of the trajectory where breakdown is most probable.

To study the degree of non-orthogonality of valence and conduction band states at different momentum points, we plot the direction and amplitude of the maximal gradient $\<u_{\mathrm{v}}(p)| \nabla_p |u_{\mathrm{c}}(p)\>$ with blue vectors in Fig.~\ref{fig:Explanation}b.  The potential breakdown region corresponds to large gradients connecting the two Fermi surfaces (as illustrated by green arrows in Fig.~\ref{fig:Explanation}b).
The second aspect, which appears to be crucial for the magnetic breakdown, is related to the value and the sign of Berry curvature.  This can be related to a notable decrease in the momentum-space velocity of the wave-packet as it passes an area of large Berry curvature. Indeed, the  semiclassical equations of motion for an electron wave-packet in band $n$ are \cite{xiao10,niu99}:
\begin{eqnarray}
\dot {\bf p} &=& \frac{-e [{\bf v} \times {\bf n_z}] B_z - e {\bf E}}{1+ e \Omega B_z/\hbar}  \label{pdot}\\
\dot {\bf r} &=& \frac{{\bf v}  + e {\bf E} \times {\bf n_z} \Omega}{1 + e \Omega  B_z/\hbar} \label{rdot}
\end{eqnarray}
where ${\bf v}(p) = {\bf \nabla} \epsilon(p) $ and the energy of the wave-packet is offset by the orbital magnetic moment.
As seen in Fig.~\ref{fig:Explanation}b, there is a large Berry curvature and small velocity near the two points of every valence band Fermi contour (these points are near the saddle point of the dispersion, similarly to the discussion in \cite{glazman18}), while the conduction band Fermi contours have higher velocity and are located at low Berry curvature regions.
Looking at the sign of Berry curvature of the valence band, we see that electrons in the valence band have higher probability to be in the breakdown-prone region for the $K$ valley due to lower momentum space velocity (larger denominator in Eq.~(\ref{pdot}) due to $e \Omega^{\mathrm{v}} B_z/\hbar > 0$).
This qualitatively explains the difference of breakdown patterns in the two valleys.
Empirical evidence from numerical studies of LL spectra at different layer numbers and $\Delta$ indicate that magnetic breakdown starts when $e \Omega^{\mathrm{v}} B_z/\hbar \gtrsim 0.6$ at the valence band Fermi line near the breakdown region indicated with green arrows.
Despite many years of studies~\cite{glazman18}, we are not aware of any similar cases studied before.
Since Eqs.~(\ref{pdot}, \ref{rdot}) were derived in the limit $|e \Omega B_z/\hbar| \ll 1$, which is no longer valid near the breakdown region, further theoretical study of this phenomemon is warranted.

\vspace{0.5 cm}
\noindent
{\bf Data availability}\newline
The data that support the findings of this study are available from the corresponding
author upon reasonable request.

\vspace{0.5 cm}
\noindent
{\bf Acknowledgments}\newline
We thank Servet Ozdemir, Vladimir Enaldiev, Yanmeng Shi, Jun Yin, and Artem Mishchenko for useful discussions. We acknowledge support from EU Graphene Flagship Project, EPSRC grants EP/S019367/1, EP/P026850/1 and EP/N010345/1, EC Project 2D-SIPC, and ERC Synergy Grant Hetero2D. M.K. acknowledges the financial support of JSPS KAKENHI Grant Number JP17K05496.

\vspace{0.5cm}

\noindent
{\bf Author contributions}\newline
S.S., E.M., M.K.\ and V.I.F.\ contributed to the formulation of the problem and development of the theory. S.S.\ performed the numerical calculations and prepared the figures. S.S., E.M.\ and V.I.F.\ wrote the manuscript.

\vspace{0.5cm}

\noindent{\bf Competing interests}\newline
The authors declare no competing interests.

\end{document}